# Atomic-scale structural fluctuations of a plasmonic cavity

Anna Rosławska[1,2,*], Pablo Merino[1,3,4], Abhishek Grewal[1], Christopher C. Leon[1], Klaus Kuhnke[1,*], Klaus Kern[1,5]

[1] Max-Planck-Institut für Festkörperforschung, D-70569, Stuttgart, Germany.

[2] Université de Strasbourg, CNRS, IPCMS, UMR 7504, F-67000 Strasbourg, France.

[3] Instituto de Ciencia de Materiales de Madrid, CSIC, E-28049, Madrid, Spain.

[4] Instituto de Física Fundamental, CSIC, E-28006, Madrid, Spain.

[5] Institut de Physique, École Polytechnique Fédérale de Lausanne, CH-1015 Lausanne, Switzerland.

* roslawska@ipcms.unistra.fr

* k.kuhnke@fkf.mpg.de

## Abstract

We study fluctuations in plasmonic electroluminescence at the single-atom limit profiting from the precision of a low-temperature scanning tunneling microscope. First, we investigate the influence of a controlled single-atom transfer on the plasmonic properties of the junction. Next, we form a well-defined atomic contact of several quanta of conductance. In contact, we observe changes of the electroluminescence intensity that can be assigned to spontaneous modifications of electronic conductance, plasmonic excitation and optical antenna properties all originating from minute atomic rearrangements at or near the contact. The observations are relevant for the understanding of processes leading to spontaneous intensity variations in plasmon-enhanced atomic-scale spectroscopies.

**Main text**

Modern nanoscale spectroscopies routinely reach single-molecule sensitivity [1,2] and are even capable of achieving contrast within individual molecules [3–5]. These methods, such as surface-enhanced Raman spectroscopy (SERS) [6], tip-enhanced Raman spectroscopy (TERS) [1,3,4,6,7], tip-enhanced photoluminescence (TEPL) [8] and scanning tunneling microscopy induced luminescence (STML) [2,5,9–15], owe their excellent resolution to the local enhancement of the electromagnetic field at a local hotspot of a metallic film, a nanoparticle or a sharp tip. Such enhancement originates from the lightning rod effect amplified by collective oscillations of charges, i.e. plasmons. Both phenomena bolster the coupling of local electromagnetic fields to the far-field, which results in an increased signal detected at a macroscopic distance from the investigated structure. Because the cavity geometry is intrinsically of atomic scale, the atomic arrangement and stability of the plasmonic antenna can play a crucial role in the enhancement mechanism. In the extreme case, the electromagnetic field is confined below a single atom, constituting a so-called picocavity [16], which enables addressing optical signals with sub-molecular resolution [11,17]. The role of the atomic structure becomes apparent also in the blinking signal attributed to atomic-scale fluctuations of the plasmonic cavity [6,18] irrespective of additional chemical or adsorption site modifications of the investigated system. The significance of both the static and fluctuating atomic structure on plasmonic properties has been studied both theoretically [19–28] and experimentally [8,18,29–32], the latter, however lacking precise characterization at the single-atom level.

Here, we address this issue using the well-controlled atomic-scale environment in a low-temperature scanning tunneling microscope (STM) in ultra-high vacuum (UHV). We build the plasmonic structures of interest by depositing single Au atoms from a Au tip on a clean Au(111) surface and approaching them again with the tip until a single-atom contact is formed. Applying a voltage bias across such a contact results in electroluminescence due to the decay of the plasmonic modes excited in the junction by the current. The resulting light emission signal can be temporally monitored [10,12,33–37]. We observe irreversible changes in the plasmonic properties of the junction as well as light intensity fluctuations on the temporal scale of seconds, both of which are correlated with the transport properties of a single-atom contact. Our results show that these modifications in the observed signal arise from minute changes in the atomic structure at or near the junction.

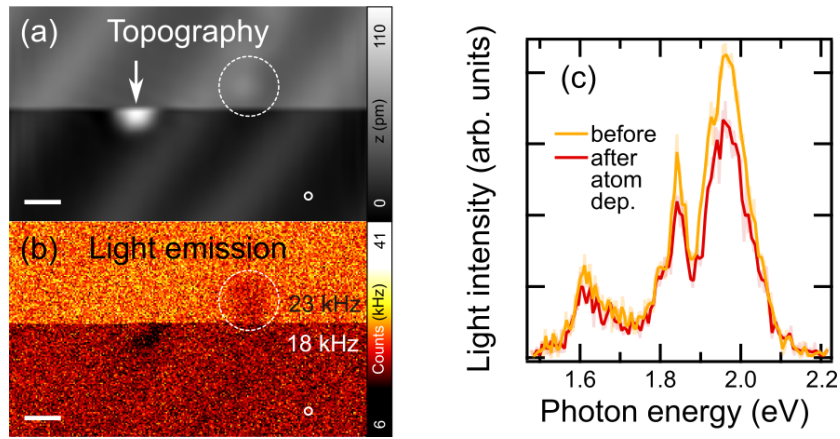

***Figure 1.*** *(a) STM topographic image of the Au(111) surface recorded under electron tunneling conditions, U = -2.5 V, I = 1 nA . During the scan (from top to bottom), a single atom was deliberately deposited from the tip apex onto the surface by atomic manipulation at the position marked by the arrow (for details see text). Scale bar: 1 nm. (b) Light intensity map recorded simultaneously with (a). The values in the bottom and upper part of the image indicate the average light intensity before and after tip modification. (c) Optical spectra recorded on the position marked by the small circle in (a,b) before (yellow curve) and after (red curve) atom deposition, U = -2.5 V,  I = 1 nA, integration time: 50 s.*

The experiments have been performed in a home-built low-temperature (4 K), UHV STM with optical access [38]. We couple its outputs to a single-photon avalanche photodiode (SPAD, MPD-PDM-R) and an optical spectrograph (Acton SP 300i, CCD camera: PI-MAX 4). The spectra presented in this work have not been corrected for the detector sensitivity. The Au(111) crystal is cleaned by repeated cycles of $Ar^+$ sputtering and annealing (up to 850 K). We used electrochemically etched [39] Au wires for tips.

First, we study the influence of a single-atom transfer on the plasmonic properties of the junction. Fig. 1(a) shows an STM image recorded from top to bottom in constant current mode. During acquisition, the scanning is stopped at the position marked by the arrow, where a single atom is deposited by closely approaching the surface with the tip (see the Supplementary Information for procedural details) [8,40]. The scan is then continued as before. The spectrally integrated light intensity is simultaneously recorded during the scan, Fig. 1(b), revealing that the deposition of an atom changes the observed signal. In the lower part of Fig. 1(b), the intensity averaged over an area of homogenous electroluminescence is 22 % lower than in the upper part for the same tunnel current. However, the shape of the spectra obtained before and after atom deposition (Fig. 1(c)) are similar, suggesting that the excited plasmonic modes in the STM junction do not change. This result shows that the electroluminescence intensity and electroluminescence spectrum can be affected differently by modifications of the tip apex. Spatially, we observe a reduced electroluminescence intensity (Fig. 1(b)) on top of the deposited atom and on a surface defect (marked by a dashed circle), which can be assigned to the local variation of the density of electronic states of the sample that modulates the electroluminescence [13,14]. However, the overall reduction of the plasmonic signal can be interpreted as a change of a local cavity structure due to a single-atom transfer, which affected both the plasmonic enhancement and the local density of states of the tip.

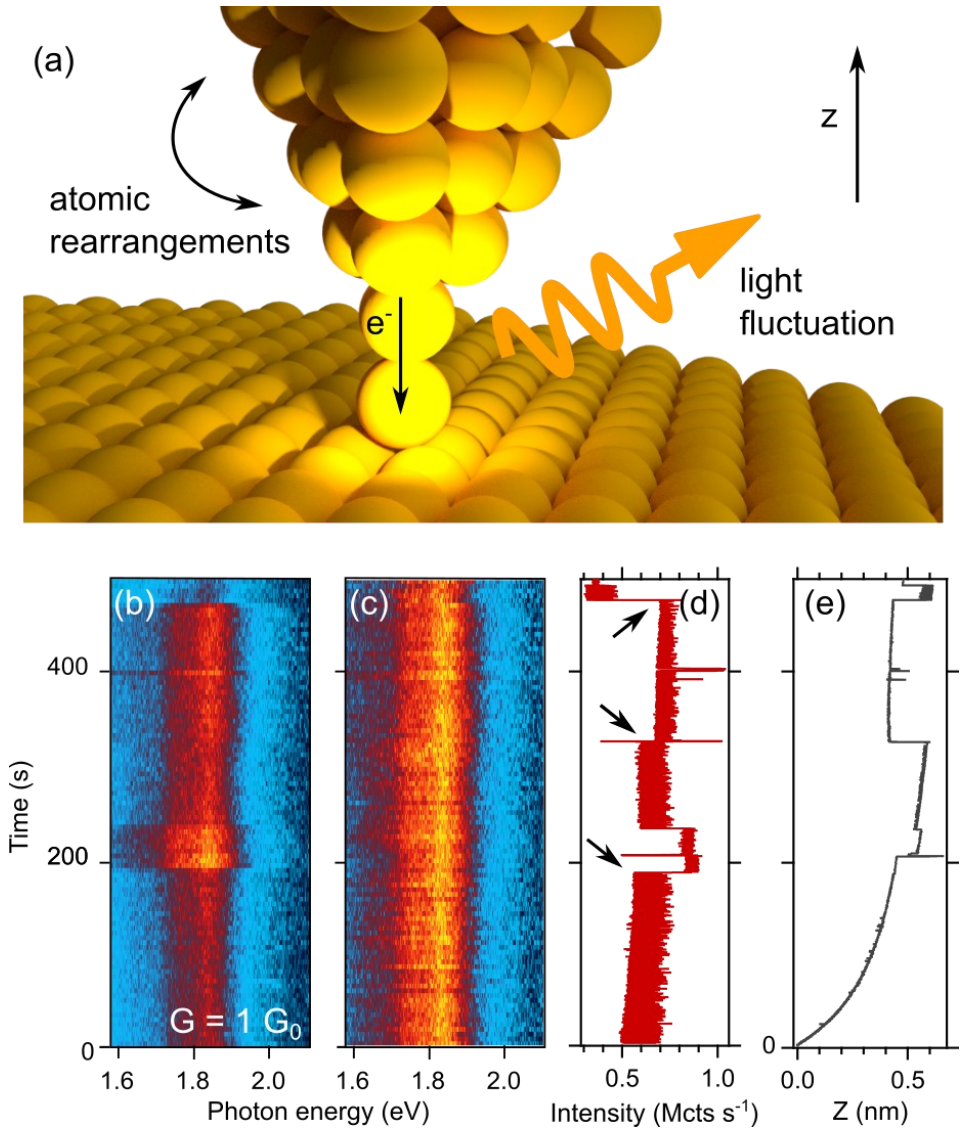

**Figure 2.** *(a) Illustration of the experiment in which the tip of an STM forms a single gold atom contact. The current passing through the junction excites the luminescence. During measurements, the current, position and light emission are monitored. (b) Time-trace of optical overbias emission spectra measured for a single-atom contact with a conductance of 1 $G_0$. The plot consists of 100 spectra; each recorded with 5 s of integration time. (c) Spectra from (b) normalized to the maximum. (d,e) Simultaneously recorded light intensity measured by the SPAD (d) and z position (e) with 20 ms integration time per point. The current feedback was enabled during the measurement to maintain 1 $G_0$, U = 1 V, I = 77.48 μA.*

We employ atomic contact experiments to increase the sensitivity for tip apex changes (Fig. 2(a)). Generally, single-atom contacts manifest their quantum nature through conductance (G) quantization. For each fully open current transmission channel, the conductance is 1 $G_0$ = 77.48 µS, as derived in quantum transport [41]. In the case of Au, a conductance of 1 $G_0$ indicates that the contact is formed between two atoms only and the conductance is dominated by one transmission channel [42]. To reduce heat dissipation at high currents in our experiment we apply a rather low voltage (on the order of 1 V) and operate the junction in an overbias light emission regime that leads to geometrically more stable junctions that undergo changes only on a time scale of seconds. Overbias emission occurs when the energy of the emitted photons exceeds the potential difference U seen by the transmitted electrons, $h\nu$ > eU [43–47]. The underlying mechanism is strongly debated in the literature and has either been assigned to plasmon-mediated coherent interaction between electrons [46,48–50] or photon emission from a hot electron gas [45,47,51].

The low-bias stability, supported by a low-temperature (4 K) environment, enables maintaining a single atom contact for several minutes using the constant current feedback loop of an STM. After a single-atom contact has been formed we turn on the feedback loop to stabilize the conductance at 1 $G_0$ by adjusting the tip height z. We then monitor as a function of time the optical spectrum (Fig. 2(b,c)), the integrated light intensity measured by the SPAD (Fig. 2(d)), and the relative change of the z position (Fig. 2(e)). During the measurement, we observe electroluminescence intensity variations while the shape of the normalized spectra (Fig. 2(c)) remains unchanged. This is remarkable since even minor voltage pulses of only a few milliseconds can modify tips enough to affect the spectral shape substantially. Such pulses can shift plasmon lines by tenths of an eV and usually affect the relative intensities of different spectral modes, but these changes are not seen here. As observed, the constant current, constant voltage condition is unable to induce such strong tip modifications.

Upon starting the measurement of Fig. 2 with some delay after settling the current, one can immediately observe a tip retraction of more than 0.5 nm (Fig. 2(e)), due to a tip elongation while the feedback loop is set to keep a constant current (77.48 μA, corresponding to 1 $G_0$) condition. The tip elongation can be linked to thermal expansion due to power dissipation (see Supplementary Material). In our study, we use relatively high bias voltages on the order of 1 V, which are necessary to drive the light emission, in contrast to bias voltages in the range of a few mV usually applied to study atomic contacts for which power dissipation effects are negligible. Remarkably, a measurement in contact results in only minor modifications to the surface, reflecting the fact that it is mostly the structure of the tip that is altered (see Supplementary Material for more details).

In addition to a continuous relaxation of geometry, rapid steps of the emission intensity can also be identified even though the light emission in Fig. 2 is driven by a constant current. With respect to the latter, we find three different types of behavior: At t = 195 s an extremely small alteration of geometry ($\Delta z$ = 6 pm) is accompanied by a significant rise of intensity (+40 %), at t = 335 s a major decrease of z ($\Delta z$ = -170 pm) leads to a small increase of intensity (+8 %), and at t = 470 s a large increase of z ($\Delta z$ = 180 pm) leads to a drastic reduction of emission (-47 %). These events are marked by arrows in Fig. 2 (d). Every such irreversible intensity modification is assigned to a geometry change at the tip or, more generally, in the junction, however, their relative size and even relative sign seem to be completely arbitrary. We also observe that fluctuations of the current due to reversible junction instabilities (noise in Fig. 2) counteracted by the feedback loop, translate into reversible fluctuations in the light emission. The results of Figure 1 and 2 together show that single atom changes to a tip apex and its repositioning on the order of a single atom length do not modify the spectral envelope of the electroluminescence but do vary its intensity drastically.

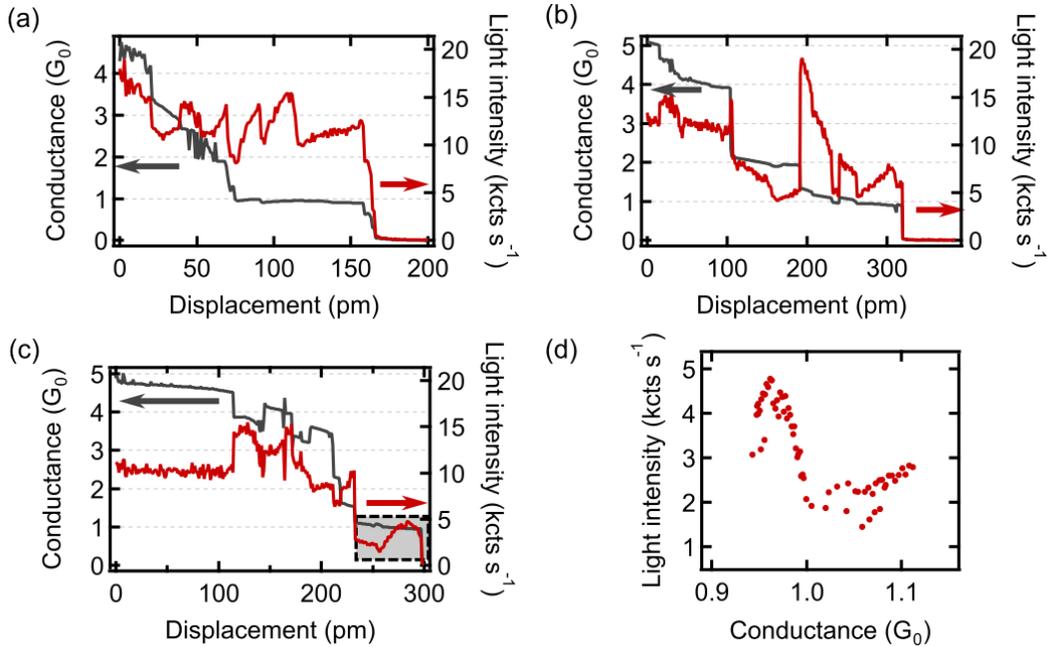

***Figure 3.*** *(a)-(c) Conductance (dark grey) and overbias light intensity (red) recorded during tip retraction with a constant speed of 5 pm s$^{-1}$ . The three different point contacts in (a)-(c) were formed by approaching the tip until a conductance of 5 $G_0$ was reached. U = -0.8 V. (d) Light intensity vs. conductance curve extracted from the data marked by the gray rectangle in (c). Note the existence of a local minimum at 1 $G_0$.*

Monitoring the geometry changes without the interference of a feedback circuit allows us to probe the relation between the optical and transport properties in more detail. We follow the rupture process of the junction by retracting the tip at a constant speed of 5 pm/s while simultaneously monitoring electroluminescence intensity and conductance (Fig. 3). In contrast to the experiment in Fig. 2, here the junction is deliberately put under increasing mechanical tension that provokes ongoing modifications to the junction. As expected from such a break-junction experiment [41], we observe steps in the conductance which are often close to multiples of $G_0$. Note that the total displacements in Figs. 3(a-c) are only 200 pm to 400 pm which is no more than 1 to 2 times the nearest neighbor distance of two Au atoms. Apart from large conductance steps, which can be attributed to a reduction of transport channels resulting from the removal of atoms from the narrowest constriction, fractional steps in the conductance are also observed. Again, conductance and light intensity changes occur together but with no correlation in relative magnitude and sign. Under these experimental conditions, minimal changes of conductance can have strong consequences for emission intensity (Fig. 3(b)). The optical spectra recorded at high conductance values (see the Supplementary Material) reveal the expected and reversible blue-shift of the plasmonic modes due to the modifications in local charge density distribution [19,24,52]. However, the result remains broadly consistent with changes at the very tip apex being primarily expressed in the electroluminescence intensity and not the electroluminescence spectral envelope. These observations point at the occurrence of a number of different processes.

Our results show that a junction operated under steady-state or dynamic conditions will incur electrical, mechanical, and thermal fluctuations that modify its electroluminescent properties. In general, the rate of light emission (R) excited by the current in an STM junction can be expressed as [46,53]

$$R(h\nu, U) = \alpha P(h\nu) T(h\nu, U) \tag{1}$$

With P(hν) being the spectral plasmonic enhancement factor depending on the optical density of states [53], T(hν, U) dictating the energy- and bias-dependent charge transport, which excites the plasmonic modes in the junction and $\alpha$ being an experimental scaling parameter. Both P(hν) and T(hν, U) depend on the geometry of the junction, which affects the spectral shape through the local plasmonic density of states and electronic transport, respectively.

First, we consider the effect of the current on the emission. At higher currents, more photons can be emitted by increasing T(hν, U). Upon decreasing conductance, we expect a more than proportional reduction of light emission due to the higher-order overbias emission mechanism. Examples can be seen in Fig. 3(a) at z = 25 pm, z = 70 pm, z = 140 pm where a conductance decrease leads to a decrease in emission. Focusing on the intensity behavior around 1 $G_0$ conductance (Fig. 3(d)) we find an emission minimum due to shot noise minimization at integer multiples of $G_0$ which has been shown to result in a minimum of light emission [44]. Such minima are often obscured in the experiment because more than one conduction channel may be involved in the charge transport.

Next, we consider the apical atoms that form the tip antenna and their reorganization due to mechanical rupture-induced processes. Comparing Figs. 3(a) and 3(b) suggests that while in (a) conductance steps reduce emission but the slow relaxation tends preferentially towards higher values of electroluminescence, in (b) the tip jumps to higher emission but relaxes mostly towards lower values of light intensity. As in the discussion of the steps in Fig. 2 also for the relaxation processes in Fig. 3, there appears to be no preferred sign of change in emission when the conductance reduces or increases. Conductance changes can be related to modifications in the atomic structure of the junction [19,41,42,47]. Variations on the order of 1 $G_0$ can be assigned to a removal/addition of single atoms at the narrowest constriction [41,54]. Finer changes may occur due to atomic rearrangements near the junction, such as an atom at the tip moving while breaking and forming bonds with its neighbors [19,42]. Rearrangements influence the transmission coefficients of the channels involved in the transport ($T(h\nu,U)$) [41], and as a consequence the luminescence changes. Similar effects can occur when the tip is expanding thermally, as in Fig. 2 (d,e). In addition to mechanical rupture due to either tip displacement or thermal effects, the atomic structure may spontaneously rearrange as a result of the current flow that induces electromigration [55], which is efficient for gold and has often been employed to fabricate single-atom junctions [47].

The changes in the plasmonic properties of the junction described by $P(h\nu)$ in eq. (1), are also related to the modifications of atomic structure [8,16,19,20,22–26,56]. In particular, Rossi [19] *et al.* and Marchesin [24] *et al.* calculated that in point contacts the plasmonic mode intensity is sensitive to the atomic-scale structure and the modifications of the modes coincide with changes in the conductance with the sign depending on the mode. While we do not observe significant changes in the emission spectra, our results agree remarkably well with these theoretical predictions regarding the total emitted intensity and confirm the occurrence of simultaneous conductance and luminescence steps. The impact on the charge flow in a slightly modified atomic structure, regardless of the plasmon excitation $T(h\nu, U)$, results in the local modification of polarizability that is reflected in the intensity of the plasmonic response $P(h\nu)$. In our experiments, we monitor a higher-order overbias emission that depends on higher powers of both $P(h\nu)$ and $T(h\nu,U)$ [46] and thus, even a subtle change in the atomic structure may result in a substantial modification of the electroluminescence. This explains the significant changes in the light intensity (as at z = 190 pm in Fig. 3(b)) observed in our experiments and demonstrates that overbias electroluminescence is an intrinsically sensitive probe to changes of the atomic structure in point contacts. Indeed, occasionally, a step change in the light intensity occurs with an insignificant change in the conductance as in Fig. 3(a) at z = 115 pm or z displacement (Fig. 2(c, d), t = 180 s), which can be attributed to an atomic rearrangement further away from the narrowest constriction.

In summary, we have investigated and characterized spontaneous variations of plasmonic luminescence output in response to different current or tip displacement control parameters at the single-atom limit. This study was carried out for electroluminescence from Au single-atom contacts in UHV at cryogenic temperature in an exceptionally well-defined environment, profiting from the precision achievable by STM. We can ascribe the observed fluctuations to current changes, as well as to mechanical stress and thermal effects that modify the atomic structure of the tip apex and reproduce key theoretical predictions [19,24]. Even under well-controlled conditions, we observe a large variety of modifications of the intrinsic plasmonic response of the system $P(h\nu)$ and the plasmon excitation efficiency $T(h\nu, U)$. They have relevance also for methods other than STML that leverage plasmonic enhancement such as SERS, TEPL and TERS. These atomic-scale phenomena are in particular critical for picocavities [8,11,16,17] and measurements in the contact-regime [7,43,57]. When local optical enhancement at the junction is considered, single-atom manipulation and stabilization of the junction may be employed to stabilize the output signal [8]. We observe the effects of slightly elevated temperature (ca. 50 K). At higher temperatures (e.g., room temperature) or under laser illumination, spontaneous atomic rearrangements are even more probable and thus result in undesired intensity blinking that would modulate the desired optical signal during investigations of more complicated systems. Such fluctuations may also affect the efficiency and stability of optical antenna devices based on the emission of plasmonic light excited by inelastic tunneling [47,58–60]. In this respect, further investigations and development of stabilization strategies will be of significant interest.

**Acknowledgements**

We would like to thank G. Schull for fruitful discussions. A.R. acknowledges support from the European Research Council (ERC) under the European Union's Horizon 2020 research and innovation program (grant agreement No 771850) and the European Union's Horizon 2020 research and innovation programme under the Marie Sklodowska-Curie grant agreement No 894434. P.M. acknowledges support from the A.v. Humboldt Foundation, the ERC Synergy Program (grant no. ERC-2013-SYG-610256, Nanocosmos), Spanish MINECO(MAT2017-85089-C2-1-R) and the "Comunidad de Madrid" for its support to the FotoArt-CM Project S2018/NMT-4367 through the Program of R&D activities between research groups in Technologies 2013, cofinanced by European Structural Funds.

# References

[1] R. Zhang, Y. Zhang, Z. C. Dong, S. Jiang, C. Zhang, L. G. Chen, L. Zhang, Y. Liao, J. Aizpurua, Y. Luo, J. L. Yang, and J. G. Hou, *Chemical Mapping of a Single Molecule by Plasmon-Enhanced Raman Scattering*, Nature **498**, 82 (2013).

[2] X. H. Qiu, G. V. Nazin, and W. Ho, *Vibrationally Resolved Fluorescence Excited with Submolecular Precision*, Science **299**, 542 (2003).

[3] J. Lee, K. T. Crampton, N. Tallarida, and V. A. Apkarian, *Visualizing Vibrational Normal Modes of a Single Molecule with Atomically Confined Light*, Nature **568**, 78 (2019).

[4] R. B. Jaculbia, H. Imada, K. Miwa, T. Iwasa, M. Takenaka, B. Yang, E. Kazuma, N. Hayazawa, T. Taketsugu, and Y. Kim, *Single-Molecule Resonance Raman Effect in a Plasmonic Nanocavity*, Nat. Nanotechnol. **15**, 105 (2020).

[5] B. Doppagne, M. C. Chong, E. Lorchat, S. Berciaud, M. Romeo, H. Bulou, A. Boeglin, F. Scheurer, and G. Schull, *Vibronic Spectroscopy with Submolecular Resolution from STM-Induced Electroluminescence*, Phys. Rev. Lett. **118**, 127401 (2017).

[6] A. B. Zrimsek, N. Chiang, M. Mattei, S. Zaleski, M. O. McAnally, C. T. Chapman, A.-I. Henry, G. C. Schatz, and R. P. Van Duyne, *Single-Molecule Chemistry with Surface- and Tip-Enhanced Raman Spectroscopy*, Chem. Rev. **117**, 7583 (2017).

[7] S. Liu, B. Cirera, Y. Sun, I. Hamada, M. Müller, A. Hammud, M. Wolf, and T. Kumagai, *Dramatic Enhancement of Tip-Enhanced Raman Scattering Mediated by Atomic Point Contact Formation*, Nano Lett. **20**, 5879 (2020).

[8] B. Yang, G. Chen, A. Ghafoor, Y. Zhang, Y. Zhang, Y. Zhang, Y. Luo, J. Yang, V. Sandoghdar, J. Aizpurua, Z. Dong, and J. G. Hou, *Sub-Nanometre Resolution in Single-Molecule Photoluminescence Imaging*, Nat. Photonics **14**, 693 (2020).

[9] K. Kuhnke, C. Große, P. Merino, and K. Kern, *Atomic-Scale Imaging and Spectroscopy of Electroluminescence at Molecular Interfaces*, Chem. Rev. **117**, 5174 (2017).

[10] A. Rosławska, C. C. Leon, A. Grewal, P. Merino, K. Kuhnke, and K. Kern, *Atomic-Scale Dynamics Probed by Photon Correlations*, ACS Nano **14**, 6366 (2020).

[11] B. Doppagne, T. Neuman, R. Soria-Martinez, L. E. P. López, H. Bulou, M. Romeo, S. Berciaud, F. Scheurer, J. Aizpurua, and G. Schull, *Single-Molecule Tautomerization Tracking through Space- and Time-Resolved Fluorescence Spectroscopy*, Nat. Nanotechnol. **15**, 207 (2020).

[12] L. Zhang, Y.-J. Yu, L.-G. Chen, Y. Luo, B. Yang, F.-F. Kong, G. Chen, Y. Zhang, Q. Zhang, Y. Luo, J.-L. Yang, Z.-C. Dong, and J. G. Hou, *Electrically Driven Single-Photon Emission from an Isolated Single Molecule*, Nat. Commun. **8**, 580 (2017).

[13] G. Hoffmann, T. Maroutian, and R. Berndt, *Color View of Atomic Highs and Lows in Tunneling Induced Light Emission*, Phys. Rev. Lett. **93**, 076102 (2004).

[14] K. Perronet, L. Barbier, and F. Charra, *Influence of the Au(111) Reconstruction on the Light Emission Induced by a Scanning Tunneling Microscope*, Phys. Rev. B **70**, 201405 (2004).

[15] H. Imada, K. Miwa, M. Imai-Imada, S. Kawahara, K. Kimura, and Y. Kim, *Real-Space Investigation of Energy Transfer in Heterogeneous Molecular Dimers*, Nature **538**, 364 (2016).

[16] F. Benz, M. K. Schmidt, A. Dreismann, R. Chikkaraddy, Y. Zhang, A. Demetriadou, C. Carnegie, H. Ohadi, B. de Nijs, R. Esteban, J. Aizpurua, and J. J. Baumberg, *Single-Molecule Optomechanics in "Picocavities,"* Science **354**, 726 (2016).

[17] T. Neuman, R. Esteban, D. Casanova, F. J. García-Vidal, and J. Aizpurua, *Coupling of Molecular Emitters and Plasmonic Cavities beyond the Point-Dipole Approximation*, Nano Lett. **18**, 2358 (2018).

[18]  A. Shiotari, T. Kumagai, and M. Wolf, *Tip-Enhanced Raman Spectroscopy of Graphene Nanoribbons on Au(111)*, J. Phys. Chem. C **118**, 11806 (2014).

[19]  T. P. Rossi, A. Zugarramurdi, M. J. Puska, and R. M. Nieminen, *Quantized Evolution of the Plasmonic Response in a Stretched Nanorod*, Phys. Rev. Lett. **115**, 236804 (2015).

[20]  M. Barbry, P. Koval, F. Marchesin, R. Esteban, A. G. Borisov, J. Aizpurua, and D. Sánchez-Portal, *Atomistic Near-Field Nanoplasmonics: Reaching Atomic-Scale Resolution in Nanooptics*, Nano Lett. **15**, 3410 (2015).

[21]  P. Garcia-Gonzalez, A. Varas, F. J. Garcia-Vidal, and A. Rubio, *Single-Atom Control of the Optoelectronic Response in Sub-Nanometric Cavities*, ArXiv:1903.08443 (2019).

[22]  P. Zhang, J. Feist, A. Rubio, P. García-González, and F. J. García-Vidal, *Ab Initio Nanoplasmonics: The Impact of Atomic Structure*, Phys. Rev. B **90**, 161407 (2014).

[23]  A. Varas, P. García-González, F. J. García-Vidal, and A. Rubio, *Anisotropy Effects on the Plasmonic Response of Nanoparticle Dimers*, J. Phys. Chem. Lett. **6**, 1891 (2015).

[24]  F. Marchesin, P. Koval, M. Barbry, J. Aizpurua, and D. Sánchez-Portal, *Plasmonic Response of Metallic Nanojunctions Driven by Single Atom Motion: Quantum Transport Revealed in Optics*, ACS Photonics **3**, 269 (2016).

[25]  T. P. Rossi, T. Shegai, P. Erhart, and T. J. Antosiewicz, *Strong Plasmon-Molecule Coupling at the Nanoscale Revealed by First-Principles Modeling*, Nat. Commun. **10**, 1 (2019).

[26]  M. Urbieta, M. Barbry, Y. Zhang, P. Koval, D. Sánchez-Portal, N. Zabala, and J. Aizpurua, *Atomic-Scale Lightning Rod Effect in Plasmonic Picocavities: A Classical View to a Quantum Effect*, ACS Nano **12**, 585 (2018).

[27]  J. M. Fitzgerald, S. Azadi, and V. Giannini, *Quantum Plasmonic Nanoantennas*, Phys. Rev. B **95**, 235414 (2017).

[28]  X. Chen, P. Liu, and L. Jensen, *Atomistic Electrodynamics Simulations of Plasmonic Nanoparticles*, J. Phys. Appl. Phys. **52**, 363002 (2019).

[29]  A. Emboras, J. Niegemann, P. Ma, C. Haffner, A. Pedersen, M. Luisier, C. Hafner, T. Schimmel, and J. Leuthold, *Atomic Scale Plasmonic Switch*, Nano Lett. **16**, 709 (2016).

[30]  P. C. Andersen, M. L. Jacobson, and K. L. Rowlen, *Flashy Silver Nanoparticles*, J. Phys. Chem. B **108**, 2148 (2004).

[31]  J. A. Scholl, A. García-Etxarri, A. L. Koh, and J. A. Dionne, *Observation of Quantum Tunneling between Two Plasmonic Nanoparticles*, Nano Lett. **13**, 564 (2013).

[32]  J. A. Scholl, A. L. Koh, and J. A. Dionne, *Quantum Plasmon Resonances of Individual Metallic Nanoparticles*, Nature **483**, 421 (2012).

[33]  A. Rosławska, P. Merino, C. Große, C. C. Leon, O. Gunnarsson, M. Etzkorn, K. Kuhnke, and K. Kern, *Single Charge and Exciton Dynamics Probed by Molecular-Scale-Induced Electroluminescence*, Nano Lett. **18**, 4001 (2018).

[34]  C. C. Leon, A. Rosławska, A. Grewal, O. Gunnarsson, K. Kuhnke, and K. Kern, *Photon Superbunching from a Generic Tunnel Junction*, Sci. Adv. **5**, eaav4986 (2019).

[35]  P. Merino, C. Große, A. Rosławska, K. Kuhnke, and K. Kern, *Exciton Dynamics of $C_{60}$-Based Single-Photon Emitters Explored by Hanbury Brown-Twiss Scanning Tunnelling Microscopy*, Nat. Commun. **6**, 8461 (2015).

[36]  P. Merino, A. Rosławska, C. C. Leon, A. Grewal, C. Große, C. González, K. Kuhnke, and K. Kern, *A Single Hydrogen Molecule as an Intensity Chopper in an Electrically Driven Plasmonic Nanocavity*, Nano Lett. **19**, 235 (2019).

[37]  Y. Luo, G. Chen, Y. Zhang, L. Zhang, Y. Yu, F. Kong, X. Tian, Y. Zhang, C. Shan, Y. Luo, J. Yang, V. Sandoghdar, Z. Dong, and J. G. Hou, *Electrically Driven Single-Photon Superradiance from Molecular Chains in a Plasmonic Nanocavity*, Phys. Rev. Lett. **122**, 233901 (2019).

[38] K. Kuhnke, A. Kabakchiev, W. Stiepany, F. Zinser, R. Vogelgesang, and K. Kern, *Versatile Optical Access to the Tunnel Gap in a Low-Temperature Scanning Tunneling Microscope*, Rev. Sci. Instrum. **81**, 113102 (2010).

[39] B. Ren, G. Picardi, and B. Pettinger, *Preparation of Gold Tips Suitable for Tip-Enhanced Raman Spectroscopy and Light Emission by Electrochemical Etching*, Rev. Sci. Instrum. **75**, 837 (2004).

[40] L. Limot, J. Kröger, R. Berndt, A. Garcia-Lekue, and W. A. Hofer, *Atom Transfer and Single-Adatom Contacts*, Phys. Rev. Lett. **94**, 126102 (2005).

[41] N. Agraït, A. L. Yeyati, and J. M. van Ruitenbeek, *Quantum Properties of Atomic-Sized Conductors*, Phys. Rep. **377**, 81 (2003).

[42] M. Dreher, F. Pauly, J. Heurich, J. C. Cuevas, E. Scheer, and P. Nielaba, *Structure and Conductance Histogram of Atomic-Sized Au Contacts*, Phys. Rev. B **72**, 075435 (2005).

[43] G. Schull, N. Néel, P. Johansson, and R. Berndt, *Electron-Plasmon and Electron-Electron Interactions at a Single Atom Contact*, Phys. Rev. Lett. **102**, (2009).

[44] N. L. Schneider, G. Schull, and R. Berndt, *Optical Probe of Quantum Shot-Noise Reduction at a Single-Atom Contact*, Phys Rev Lett **105**, (2010).

[45] A. Downes, Ph. Dumas, and M. E. Welland, *Measurement of High Electron Temperatures in Single Atom Metal Point Contacts by Light Emission*, Appl. Phys. Lett. **81**, 1252 (2002).

[46] P.-J. Peters, F. Xu, K. Kaasbjerg, G. Rastelli, W. Belzig, and R. Berndt, *Quantum Coherent Multielectron Processes in an Atomic Scale Contact*, Phys. Rev. Lett. **119**, 066803 (2017).

[47] M. Buret, A. V. Uskov, J. Dellinger, N. Cazier, M.-M. Mennemanteuil, J. Berthelot, I. V. Smetanin, I. E. Protsenko, G. Colas-des-Francs, and A. Bouhelier, *Spontaneous Hot-Electron Light Emission from Electron-Fed Optical Antennas*, Nano Lett. **15**, 5811 (2015).

[48] F. Xu, C. Holmqvist, and W. Belzig, *Overbias Light Emission Due to Higher-Order Quantum Noise in a Tunnel Junction*, Phys. Rev. Lett. **113**, (2014).

[49] F. Xu, C. Holmqvist, G. Rastelli, and W. Belzig, *Dynamical Coulomb Blockade Theory of Plasmon-Mediated Light Emission from a Tunnel Junction*, Phys. Rev. B **94**, 245111 (2016).

[50] K. Kaasbjerg and A. Nitzan, *Theory of Light Emission from Quantum Noise in Plasmonic Contacts: Above-Threshold Emission from Higher-Order Electron-Plasmon Scattering*, Phys. Rev. Lett. **114**, (2015).

[51] M. Buret, I. V. Smetanin, A. V. Uskov, G. Colas des Francs, and A. Bouhelier, *Effect of Quantized Conductivity on the Anomalous Photon Emission Radiated from Atomic-Size Point Contacts*, Nanophotonics **0**, (2019).

[52] R. Esteban, A. G. Borisov, P. Nordlander, and J. Aizpurua, *Bridging Quantum and Classical Plasmonics with a Quantum-Corrected Model*, Nat. Commun. **3**, 825 (2012).

[53] M. Parzefall and L. Novotny, *Light at the End of the Tunnel*, ACS Photonics **5**, 4195 (2018).

[54] H. Ohnishi, Y. Kondo, and K. Takayanagi, *Quantized Conductance through Individual Rows of Suspended Gold Atoms*, Nature **395**, 6704 (1998).

[55] M. Ring, D. Weber, P. Haiber, F. Pauly, P. Nielaba, and E. Scheer, *Voltage-Induced Rearrangements in Atomic-Size Contacts*, Nano Lett. **20**, 5773 (2020).

[56] X. Chen and L. Jensen, *Morphology Dependent Near-Field Response in Atomistic Plasmonic Nanocavities*, Nanoscale **10**, 11410 (2018).

[57] G. Reecht, F. Scheurer, V. Speisser, Y. J. Dappe, F. Mathevet, and G. Schull, *Electroluminescence of a Polythiophene Molecular Wire Suspended between a Metallic Surface and the Tip of a Scanning Tunneling Microscope*, Phys. Rev. Lett. **112**, (2014).

[58] M. Parzefall and L. Novotny, *Optical Antennas Driven by Quantum Tunneling: A Key Issues Review*, Rep. Prog. Phys. **82**, 112401 (2019).

[59] J. Kern, R. Kullock, J. Prangsma, M. Emmerling, M. Kamp, and B. Hecht, *Electrically Driven Optical Antennas*, Nat. Photonics **9**, 582 (2015).

[60] F. Bigourdan, J.-P. Hugonin, F. Marquier, C. Sauvan, and J.-J. Greffet, *Nanoantenna for Electrical Generation of Surface Plasmon Polaritons*, Phys. Rev. Lett. **116**, 106803 (2016).

Supplementary material for:

## Atomic-scale structural fluctuations of a plasmonic cavity

Anna Rosławska[1,2,*], Pablo Merino[1,3,4], Abhishek Grewal[1], Christopher C. Leon[1], Klaus Kuhnke[1,*], Klaus Kern[1,5]

[1] Max-Planck-Institut für Festkörperforschung, D-70569, Stuttgart, Germany.

[2] Université de Strasbourg, CNRS, IPCMS, UMR 7504, F-67000 Strasbourg, France.

[3] Instituto de Ciencia de Materiales de Madrid, CSIC, E-28049, Madrid, Spain.

[4] Instituto de Física Fundamental, CSIC, E-28006, Madrid, Spain.

[5] Institut de Physique, École Polytechnique Fédérale de Lausanne, CH-1015 Lausanne, Switzerland.

* roslawska@ipcms.unistra.fr

* k.kuhnke@fkf.mpg.de

## 1. Single-atom deposition

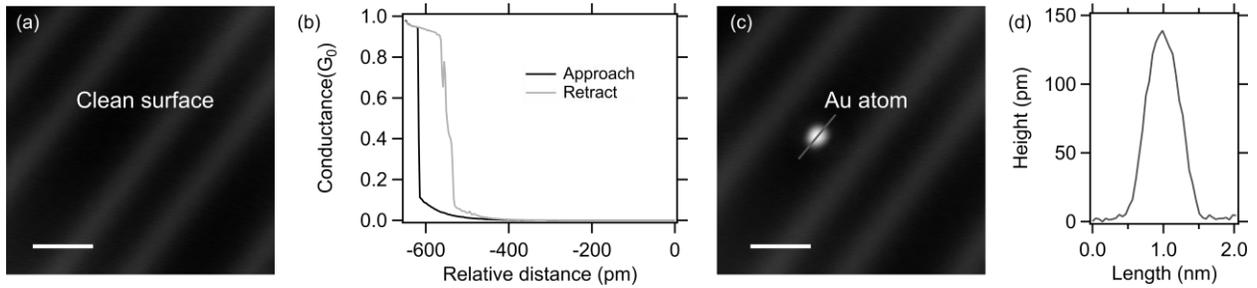

**Figure S1.** *Deposition of individual Au atoms. a) Image of a defect-free Au(111) surface, I = 100 pA, U = 1 V. b) Approach and retract curve resulting in deposition of a single atom, U = 1 V, current set-point 100 pA. c) Scan of the same area as in a) after atom deposition. Scale bars 2 nm. d) Height profile showing a cross-section (marked in c)) over the deposited atom.*

We prepare atomic-scale contacts used in our study by depositing individual atoms on the Au(111) surface. It can be controllably achieved by approaching the surface with the tip until contact [1]. Fig. S1 shows the details of this procedure. First, the surface is scanned to ensure it is clean, as presented in Fig. S1(a). Next, the tip is moved towards the surface by 650 pm which results in a jump-to-contact event leading to a conductance close to 1 $G_0$ (Fig. S1(b)); the tip is subsequently retracted. In the next step, the surface is imaged again to confirm successful atomic deposition, as shown in Fig. S1(c). We find this procedure to be highly reproducible, which on occasion may also deposit a small cluster of a few atoms, which can be readily identified by a higher maximum conductance during the approach-retract curve and a higher topographic appearance than the one presented in Fig. S1(d). Since both the tip and the sample are made of Au, the deposited structures consist of Au atoms only.

**2. Measurements with the feedback loop closed (Fig. 2)**

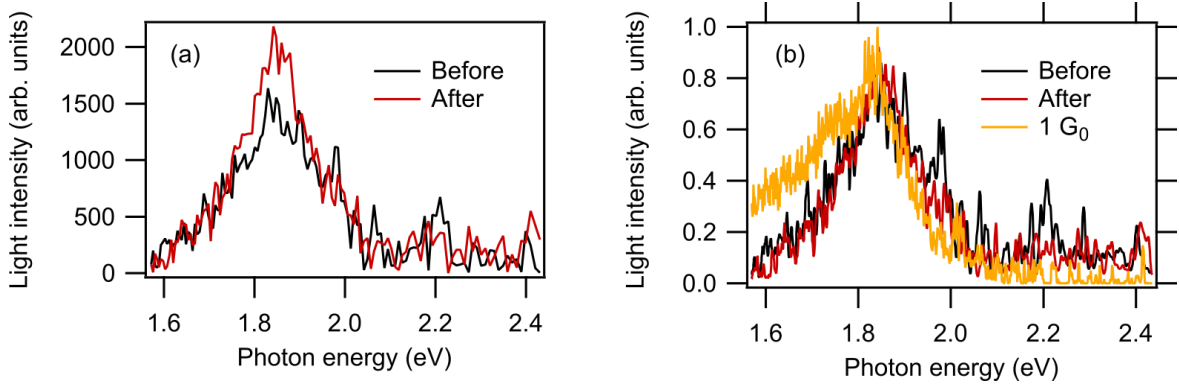

***Figure S2.*** *(a) Optical spectra (binned) recorded on a clean Au(111) surface before and after measurements shown in Fig. 2 of the main manuscript, U = 3 V, I = 100 pA, integration time: 30 s. (b) Normalized reference spectra from (a) together with one of the spectra from Fig. 2(b), U = 1 V, I = 77.48 μA.*

Fig. S2 presents optical spectra recorded, before, during and after measurements presented in Fig. 2 of the main manuscript. We find that the shape of the spectrum did not change significantly after the measurement. The total integrated intensity, however, increased by 14 % (Fig. S2(a)). In Fig. S2(b) we compare these reference spectra in tunnel conditions (100 pA) with the spectrum recorded in contact (77 μA) and overbias emission condition. The main mode is slightly red-shifted with respect to the spectra recorded in tunneling. The overall higher intensity in the low energy regime (<1.8 eV) is a result of the overbias emission, which is less efficient at higher energies.

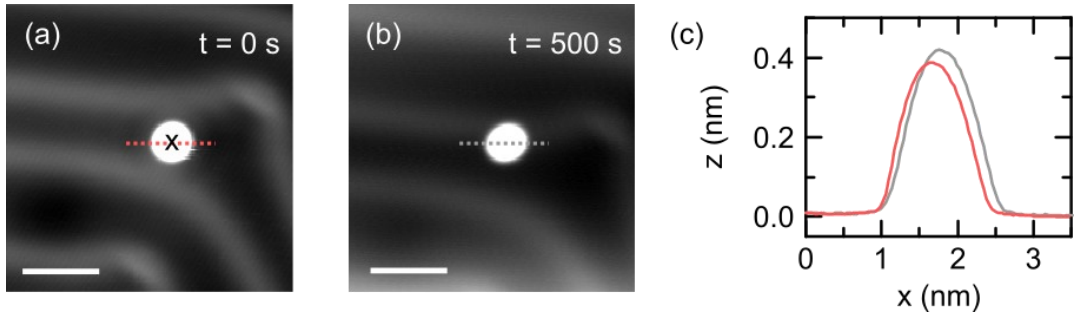

**Figure S3.** (a), (b) STM images recorded before and after the contact measurement shown in Fig.2 of the main manuscript. The contact was formed at the marked position. U = 1 V, I = 100 pA, scale bars 3 nm. (c) Height profiles measured along the lines indicated in (a) and (b).

Fig. S3 shows the area of the sample used to perform measurements presented in Fig. 2 of the main manuscript. The single-atom contact has been formed on top of the structure deposited from the tip as marked in Fig. S3(a). To evaluate the permanent modifications of the junction after the experiment, the surface was rescanned (Fig. S3(b)). We find that the investigated structure moved by 1 nm and likely an atom has been transferred to the surface, as indicated by the increased apparent height shown in Fig. S3(c). In this particular experiment, we contacted a larger adsorbed structure consisting of more than one atom because such junctions are more prone to spontaneous changes and thus more illustrative for our study. For the sake of completeness, we performed measurements in which we contacted a single atom deposited on the surface and observed both fluctuations when in contact and the permanent modification of electroluminescence as compared before and after the measurement. This experiment is described in the next section (Fig. S4 and Fig. S5).

**3. Luminescence fluctuations in a point contact to a single atom deposited on the surface**

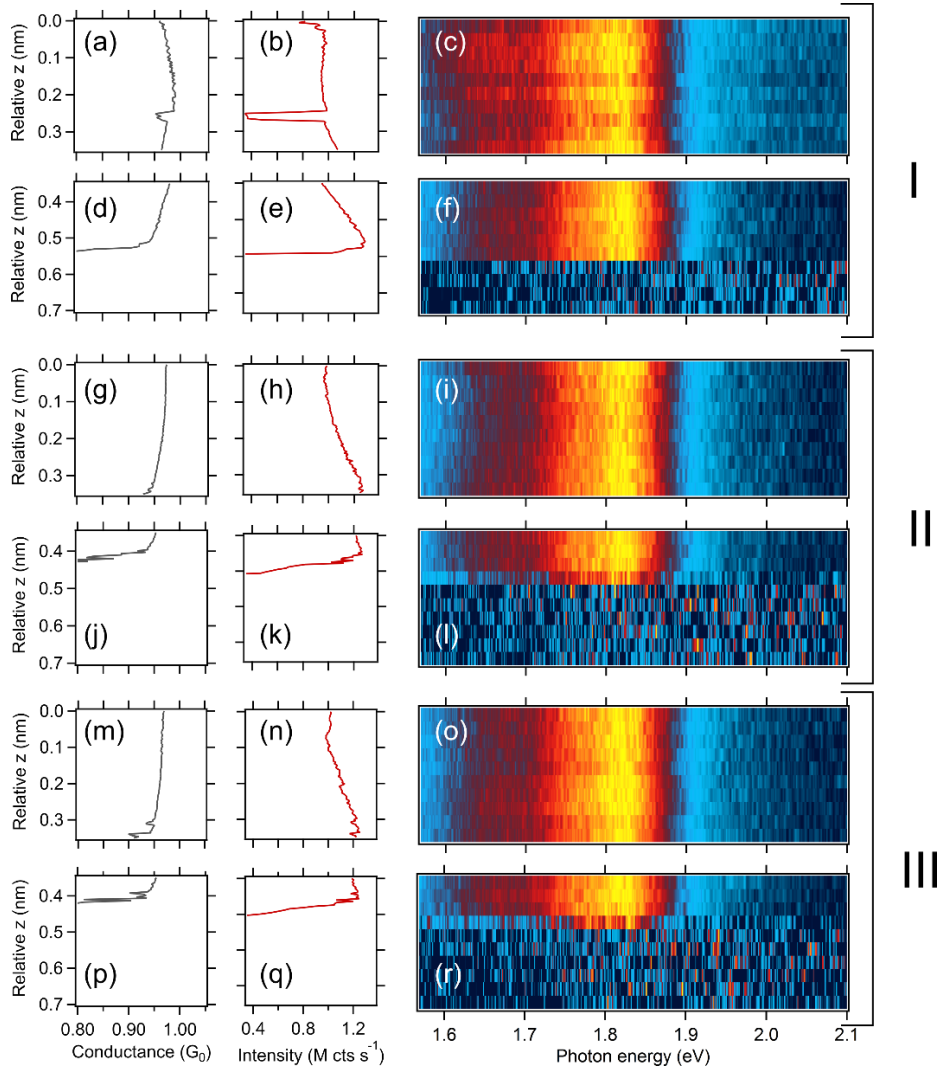

***Figure S4.*** *Conductance and luminescence fluctuations during retraction from an atomic contact until rupture. The contact was formed on a single adatom as described earlier. The left column shows conductance traces, the middle column the spectrally integrated light intensity and, the right column emission spectra normalized to their maximum. Each plot in the right column consists of 10 spectra (5 s integration time). The whole figure presents 3 separate experiments (I, II, III), each recorded in two parts (for example a)-c) and d)-f)) accounting for the start of the stretching and the breaking of contact. In each part, the tip was retracted by 350 pm in steps of 3.5 pm. Between the separate experiments, the*

*surface was inspected by a topographic scan and after that, the atomic contact was reestablished.*

*Overbias emission condition U = 1 V.*

In the experiments displayed in Fig. S4 we deposited a single Au atom (see Fig. S5), established a contact of 1 $G_0$, and then retracted the tip until loss of contact (rupture). Simultaneously, we monitored the electroluminescence integrated intensity (middle column in Fig. S4) and the overbias emission spectrum (right column in Fig. S4). No current feedback was used throughout the experiment. The measurements were performed in an extremely gentle fashion, each time the tip was continuously retracted (in steps of 3.5 pm) by 350 pm in total. If the contact did not break, the tip was retracted by another 350 pm. We found that typically the contact broke after a total retraction of approx. 450 pm. After each experiment, we rescanned the surface and repeated the experiment on the same atom following the same protocol. This procedure was reproducible and the approach-retract routine could be repeated on the same atom multiple times, as shown in Fig. S4. Similarly to the measurements presented in Fig. 2 and Fig. 3 of the main text, the luminescence can evolve in both gradual and step-wise manner which can be correlated with the changes in the conductance (left column in Fig. S4). Again, the spectral shape remains unchanged as displayed in the right column of Fig. S4. The tip was stabilized for 15 h before the series of experiments was started.

Fig. S5 shows STM images recorded before (Fig. S5(a)) and after (Fig. S5(b)) accomplishing all experiments presented in Fig. S4, the investigated atom (deposited from the tip as described above) is marked by an arrow. We find that at the end of the experiments the deposited adatom had moved by 1 nm in total but no additional atom had been transferred, which demonstrates the reproducible experimental conditions. In contrast, the apex of the tip has evidently changed: First, the deposited atom and the defects on the surface are imaged as features with smaller extension in Fig. S5(b) than in Fig. S5(a) indicating that the tip apex has become sharper. Second, there is a decrease in the light intensity by 8 % percent (Fig. S5(c)).

These observations further confirm that a minute change at the tip structure, even without transferring an atom to the surface, may induce a modification in the plasmonic properties of the junction.

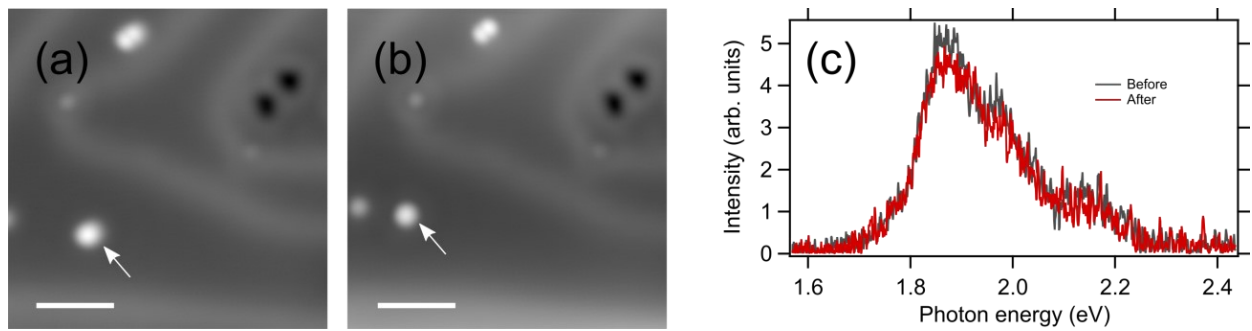

*Figure S5. Stability of a single atom on the surface during measurements in contact. (a), (b) STM images recorded before and after measurements presented in Fig. S4, which were performed on the atom marked by an arrow. U = 1 V, I = 100 pA, scale bars 2 nm. (c) Optical spectra recorded before and after the measurements presented in Fig. S4, tunnel conditions U = 3 V, I = 100 pA.*

## 4. Local heat dissipation in the junction

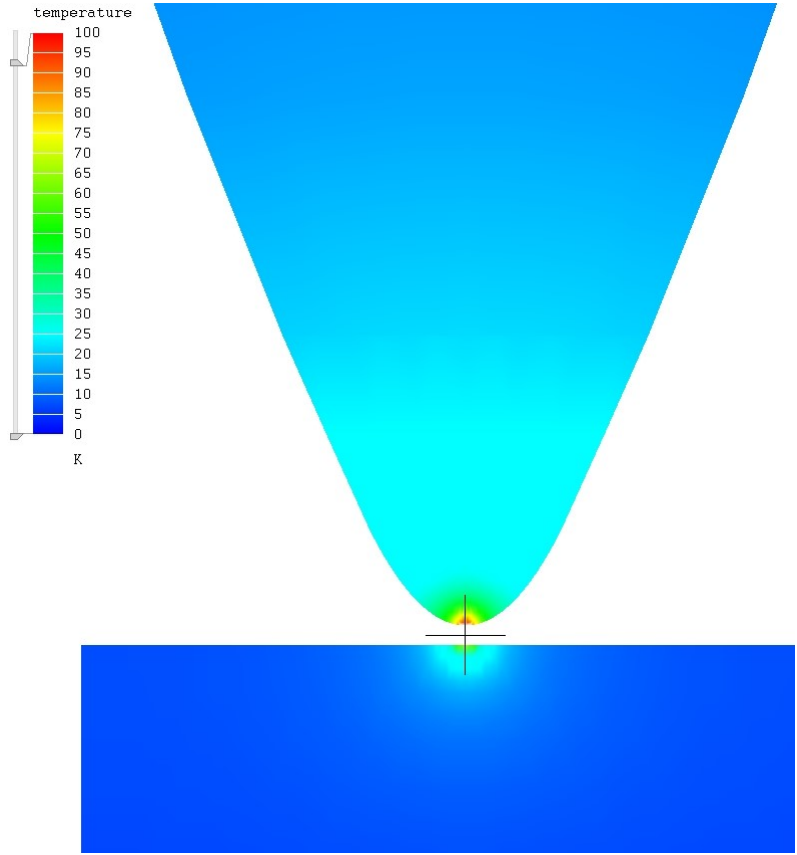

***Figure S6.*** *Temperature distribution in a plane cut through the 3-dimensional geometry of the junction. The distribution was determined using the Mecway finite element software (Mecway Ltd., New Zealand). We used the temperature-dependent thermal conductivity reported for unannealed high purity (99.99%) Au wires [2]. Tip curvature radius ca. 3 nm, the opening angle of a cone ca. 45°. Length scale: The tip-sample distance is set to 1 nm exactly. A power of 40 μW (half of the dissipation at 1 $G_0$) is input over an area of ca 1 $nm^2$ at the contact point independently to both tip and sample. No further heat transfer between tip and sample is assumed.*

In Fig. 2e of the main text, we observe thermal tip elongation due to the power dissipated in the junction. Since the main part of the applied voltage drops across the tip-substrate contact region, we can estimate the dissipated power at the junction to be close to $U^2 G_0 \approx 77$ μW. About half of the power will be dissipated on the tip and substrate each. However, since heat transport in the substrate occurs over a hemispherical ($2\pi$) region in the macroscopic Au crystal while the tip is sharp and conical, the strongest contribution to thermal expansion will be due to heat dissipation in the tip. Coarse modeling of heat flow at a conical gold tip suggests a temperature of the order of 30 K to 100 K (see Fig. S6) which may, however, be surpassed in close vicinity to the atomic junction. The simulation demonstrates that the conical tip exhibits stronger heating due to the reduced effective dimensionality in comparison to the half-space available for heat transport in the substrate.

In laser-based plasmon-enhanced spectroscopies (as e.g. TERS or TEPL) the thermal input by the illumination can become comparable to the power dissipation by the electric current and higher ambient temperatures will play an even greater role. While our study tries to limit heat input by using low voltages and cryogenic temperature, the thermal reorganization will be most relevant at higher conductance (4-5 $G_0$) since here the heat dissipation at the junction is increased by the respective prefactor.

## 5. Optical spectroscopy at high conductances

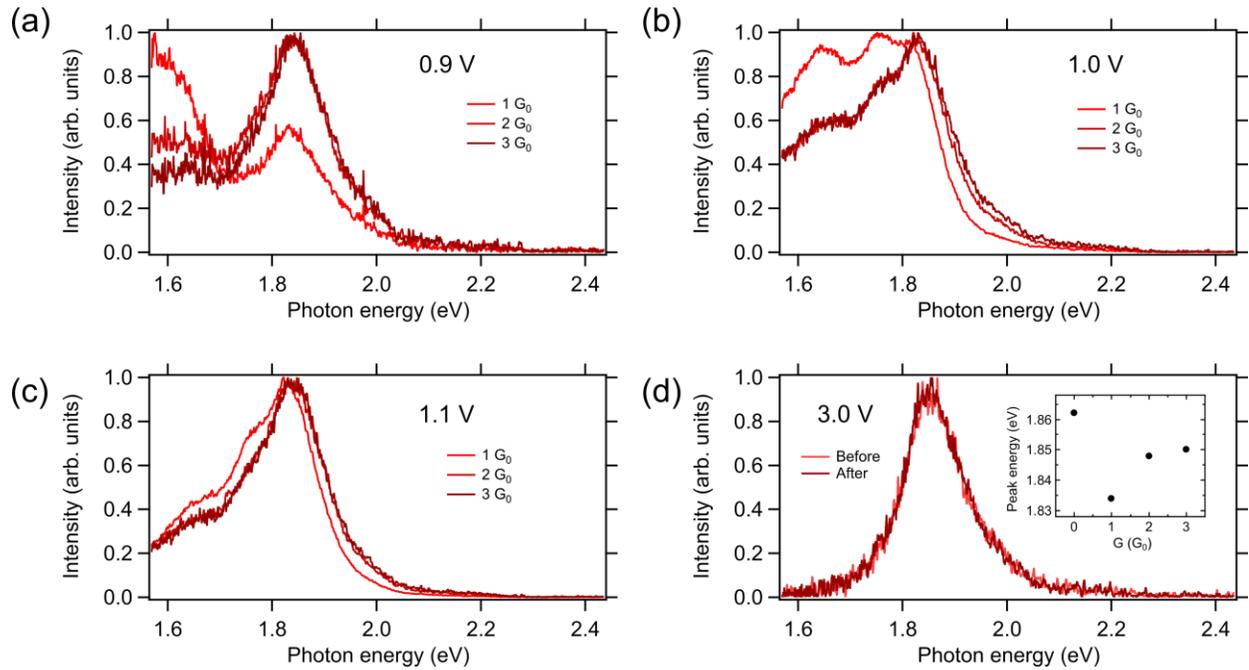

***Figure S7.*** *Optical spectra as a function of conductance and bias. (a)-(c) Normalized optical overbias spectra as a function of bias and conductance, with the values indicated next to the traces. The current feedback was enabled during the measurements. (d) Normalized reference spectra recorded at tunneling conditions before and after the measurements presented in a)-c), U = 3 V, I = 100 pA (corresponding to $4 \times 10^{-7} G_0$). The inset shows the energies of the main plasmonic mode as a function of conductance, the values for measurements in contact are taken from data in (c).*

Fig. S7 presents a series of spectra recorded for different bias voltages and different conductances. Fig. S7(d) shows reference spectra recorded before and after the series to demonstrate that the plasmonic modes of the junction have not been altered by the measurement. The shapes of the spectra are strongly bias-dependent because of the strong non-linearity of the overbias emission. Remarkably, the relative intensity of the low-energy component is highest at 1 $G_0$. For each studied bias in Fig. S7a-c, the spectra for 2 and 3 $G_0$ nearly overlap with a well-pronounced maximum around 1.8 eV, whereas at 1 $G_0$ the luminescence intensity at $h\nu < 1.8$ eV may be as high or even higher than the intensity of 1.8 eV maximum, as in Fig. S7b. This effect may be related to a reduced probability of electron-electron interaction [3] due to the limited number of participating transport channels under these conditions, which can be as low as 1. When the electrons pass mainly through one channel, one may expect the interaction between successive electrons to be reduced due to their spatiotemporal separation. We would like to note that the peak energy of the main mode shifts by about 30 mV. It is highest at tunneling conditions (1.86 eV), lowest (1.83 eV) at 1 $G_0$, and increases again with the conductance reaching 1.85 eV at 3 $G_0$ (see the inset in Fig. S7(d)). This is in qualitative agreement with the ab initio [4,5] and quantum plasmonic predictions [6].

**References**

[1] L. Limot, J. Kröger, R. Berndt, A. Garcia-Lekue, and W. A. Hofer, *Atom Transfer and Single-Adatom Contacts*, Phys. Rev. Lett. **94**, 126102 (2005).

[2] G. K. White, *The Thermal Conductivity of Gold at Low Temperatures*, Proc. Phys. Soc. Sect. A **66**, 559 (1953).

[3] P.-J. Peters, F. Xu, K. Kaasbjerg, G. Rastelli, W. Belzig, and R. Berndt, *Quantum Coherent Multielectron Processes in an Atomic Scale Contact*, Phys. Rev. Lett. **119**, 066803 (2017).

[4] T. P. Rossi, A. Zugarramurdi, M. J. Puska, and R. M. Nieminen, *Quantized Evolution of the Plasmonic Response in a Stretched Nanorod*, Phys. Rev. Lett. **115**, 236804 (2015).

[5] F. Marchesin, P. Koval, M. Barbry, J. Aizpurua, and D. Sánchez-Portal, *Plasmonic Response of Metallic Nanojunctions Driven by Single Atom Motion: Quantum Transport Revealed in Optics*, ACS Photonics **3**, 269 (2016).

[6] R. Esteban, A. G. Borisov, P. Nordlander, and J. Aizpurua, *Bridging Quantum and Classical Plasmonics with a Quantum-Corrected Model*, Nat. Commun. **3**, 825 (2012).